\documentclass[a4paper,11pt]{article}
\usepackage[utf8]{inputenc}
\usepackage[english]{babel}
\usepackage{amsmath, amssymb,graphicx}
\usepackage{braket}
\usepackage{yfonts}
\usepackage{caption}
\usepackage{subcaption}
\usepackage{ulem}
\pdfoutput=1
\usepackage{jheppub}

\title{Path Integrals in Quadratic Gravity}

\author[a,b]{Vladimir V. Belokurov}
\author[a]{and Evgeniy T. Shavgulidze}

\affiliation[a]{Lomonosov Moscow State University,}

\affiliation[b]{Institute for Nuclear
Research of the Russian Academy of Sciences, }

\emailAdd{vvbelokurov@yandex.ru}
\emailAdd{shavgulidze@bk.ru}

\abstract
{ Using the invariance of Quadratic Gravity in FLRW metric under the group of diffeomorphisms of the time coordinate,
we rewrite the action $A$ of the theory in terms of the invariant dynamical variable $g(\tau)\,.$

We propose to consider the path integrals  $\int\,F(g)\,\exp\left\{-A \right\}dg$ as the integrals over the functional measure $\mu(g)=\exp\left\{-A_{2} \right\}dg\,,\ $ where $A_{2}$ is the part of the action $A$ quadratic in $R\,.$
The rest part of the action in the exponent stands in the integrand as the "interaction" term.
We prove the measure $\mu(g)$  to be equivalent to the Wiener measure, and, as an example, calculate the averaged scale factor in the first nontrivial perturbative order.}

\keywords{Quadratic Gravity, Path integrals, Wiener measure}

\arxivnumber{.}

\begin{document}
\maketitle

\section{ Introduction}

\vspace{0.5cm}

One of the challenging problems of modern physics is the development of a quantum theory of gravity having at least the same level of reliability
as the existing theories of the other fundamental interactions.
At the first sight, it would be quite natural to consider the quantum field theory given by the Einstein-Hilbert action with the metric tensor $\mathcal{G}_{\mu\nu}$ being the quantum field. However, this way one obtains a non-renormalizable theory.

The traditional and the most straightforward solution to the problem is to add the fourth-derivative terms to the action . That is to consider
the Quadratic Gravity action of the form
\begin{equation}
   \label{ActionGen}
\mathcal{A} =\mathcal{A}_{0}+\mathcal{A}_{1}+\mathcal{A}_{2}\,,
\end{equation}
where
\begin{equation}
   \label{A0}
\mathcal{A}_{0} =\Lambda\int\,d^{4}x\,\sqrt{-\mathcal{G}}\,,
\end{equation}\begin{equation}
   \label{A1}
\mathcal{A}_{1} \equiv \mathcal{A}_{EH} =-\frac{\kappa}{6}\,\int\,d^{4}x\,\sqrt{-\mathcal{G}}\,R\,,\ \ \ \ \ \kappa>0\,,
\end{equation}
$$
\mathcal{A}_{2} =\int\,d^{4}x\,\sqrt{-\mathcal{G}}\,\left(\tilde{c}_{1}\,R^{2}+\tilde{c}_{2}\,R_{\mu\nu}R^{\mu\nu}+\tilde{c}_{3}\,R_{\mu\nu\varrho\sigma}
R^{\mu\nu\varrho\sigma}\right)
$$
\begin{equation}
   \label{A2}
 =\int\,d^{4}x\,\sqrt{-\mathcal{G}}\,\left(c_{1}\,R^{2}+c_{2}\,GB+c_{3}\,C^{2}\right)\,.
\end{equation}
Here,
$$
GB=\sqrt{-\mathcal{G}}\,\left( R^{2}-4\,R_{\mu\nu}R^{\mu\nu}+R _{\mu\nu\varrho\sigma}\,R^{\mu\nu\varrho\sigma}\right)
$$
is the the Gauss-Bonnet term, and
$$
C^{2}=C _{\mu\nu\varrho\sigma}\,C^{\mu\nu\varrho\sigma}=R _{\mu\nu\varrho\sigma}\,R^{\mu\nu\varrho\sigma}-2R_{\mu\nu}R^{\mu\nu}+\frac{1}{3}R^{2}
$$
is the square of the Weyl tensor.

The properties of quadratic gravity, including its perturbative renormalisability, the spectrum of propagating modes, its representation as the standard Einstein gravity interacting with additional fields, Newtonian limit, its role in the description of cosmological inflation, black hole solutions, and the relation with string theory, were studied intensively.
(See the discussion of the subjects, e.g., in   some early and recent papers and reviews \cite{(Utiyama)} - \cite{(Donoghue2)}.)

Unfortunately, even if the theory is formally renormarizable, generally speaking, it is not consistent because of the presence of the ghosts that are the fields with negative kinetic energy in the action.
A comprehensive review of the origins of ghosts, their classification, as well as various examples of dynamical systems with ghosts
are given in \cite{(Woodard)} - \cite{(Smilga4)}.
In general, the presence of ghosts violates the unitarity of the quantum theory of gravity and makes it unstable with respect to the general metric perturbations.
Although there are attempts to cure the theory in some special  way ( e.g., with the addition of compensating fields due to the more complicated geometry \cite{(Shapiro_torsion)} or supersymmetry \cite{(Smilga5)} - \cite{(Smilga6)}),
ghosts are considered as the main obstacle for constructing a consistent theory of quantum gravity.
See the discussion  of the problem of ghosts in higher-derivative models of gravity in the recent thoroughgoing book \cite{(Buchbinder)}.

To construct a quantum theory of gravity one can try to consider the path integrals over the space of metrics $ \mathcal{G}\,.$

 This way, however, one meets the serious problem.
The point is that even the EH action $\mathcal{A}_{1}$ is unbounded from below \cite{(Gibbons)}. The EH action is conformally equivalent to the
conformal scalar-tensor action.
Although the tensor spin-two mode is positively defined, the scalar field has a negative sign of kinetic term. On the other hand, the scalar mode itself is a gauge degree of freedom.

However, if we fix the gauge and restrict the consideration by the FLRW metric
\begin{equation}
   \label{FLRW}
ds^{2} =N^{2}(t)\,dt^{2}-a^{2}(t)\,d\vec{x}^{2}\,,\ \ \ \ \ N(t)>0\,,\ \ a(t)>0\,,
\end{equation}
the problem becomes apparent. In this case,
\begin{equation}
   \label{gR}
\sqrt{-\mathcal{G}}= N(t)\,a^{3}(t)\,,\ \ \ R(N(t),\,a(t))=-6\left(\frac{a''}{N^{2}a}+\frac{(a')^{2}}{N^{2}a^{2}}-\frac{N'\,a'}{N^{3}a} \right)\,,
\end{equation}
and  the EH action normalized to the unit space volume is written as
\begin{equation}
   \label{Action1}
A_{1}\equiv\frac{\mathcal{A}_{1}}{V_{3}} =-\kappa\,\int\,dt\,\frac{a\,(a')^{2}}{N}+\,BT
\end{equation}
with the boundary term
\begin{equation}
   \label{BT}
BT =\kappa\,\int\,dt\,\left(\frac{a^{2}a'}{N}\,\right)' \,.
\end{equation}

Therefore,
$$
\exp\left\{-\mathcal{A}_{1} \right\}\,d\mathcal{G}
$$
cannot be considered as a functional measure of Euclidean path integration.

To overcome the trouble with Euclidean path integrals in general relativity, several approaches were proposed.
(See some ideas and  results   in
\cite{(Gibbons2)} - \cite{(Narain)}). The most radical ones avoid Wick rotation and study Feynman path  integrals directly.
The lack of a countably-additive measure for functional integration is a common feature of the approaches. Therefore they deal with the integrals that are not absolutely convergent.
Note that in the ordinary quantum field theory in the Minkowski space-time,
one has the opposite situation. Euclidean path integrals are used  as a justification for Feynman path integrals in the theory.

Another idea (\cite{(LoukoSorkin)} - \cite{(Visser)}) is to use the complex space-time metrics  to validate path integrals in gravity.

In this paper, we propose the completely different approach to quantize gravity. We consider $R+R^{2}$ theory in the FLRW metric and find dynamical variable $g(\tau)$ that is invariant under the group of diffeomorphisms of the time coordinate.
Then we turn  the Feynman path integrals
\begin{equation}
   \label{FPIA}
\int\,F(g)\,\exp\left\{i\,A (g)\right\}\,dg
\end{equation}
into the Euclidean ones
\begin{equation}
   \label{EPIA1}
\int\,F(g)\,\exp\left\{-A (g)\right\}\,dg
\end{equation}
by the corresponding transformation of the space-time metric.

Let us stress that we consider path integrals not over the space of metrics $\mathcal{G}$, as it is usually done, but
over the space of continuous functions $g(\tau)$ related to the conformal factor of the metric. In some sense, it is a particular realization of the idea
of the seminal papers \cite{(Antoniadis0)},  \cite{(Antoniadis2)} to consider the effective theory of the conformal factor as the true (at least at some scales) theory of quantum gravity. (See the discussion in sections \ref{sec:inv} and \ref{sec:concl}.)

The second our novelty is to treat (\ref{EPIA1}) as the integrals over the functional measure $\mu(g)=\exp\left\{-A_{2} \right\}dg\,,$ where $A_{2}$ is the part of the action $A$ quadratic in $R\,.$ The rest part of the action in the exponent stands in the integrand as the "interaction" term.
We prove the measure $\mu(g)$  to be equivalent to the Wiener measure, and, as an example, calculate the averaged scale factor in the first nontrivial perturbative order.

First, in section  \ref{sec:inv}, we specify the problem and express the quadratic gravity action in FLRW metric in terms of the function $g$ invariant under the group of diffeomorphisms of the time coordinate.

In section \ref{sec:measure}, we construct the functional integrals measure and prove that it is equivalent to the Wiener measure.

Then, in section \ref{sec:pertcorr}, we calculate the first nontrivial perturbative corrections to the averaged scale factor $<a(t)>$.
In section \ref{sec:concl}, we summarize our results and discuss briefly the difference between our approach and the other ones. Also we give  some remarks connected with the problems for possible further studies.

Appendices  contain some auxiliary information of technical character.
In particular, in appendix A,  we find the classical solutions of the Euler-Lagrange equation.
One of them gives the power behavior of the  classical scale factor
$(a_{0}^{(1)}\sim \sqrt{t} )\,,$ and the other leads to the exponential behavior $(a_{0}^{(2)}\sim \exp\{ct \} )\,.$
In appendix B, we present the explicit form of the integrals calculated in section \ref{sec:pertcorr}.
\vspace{0.5cm}

\section{Invariance of the action and diffeomorphism invariant dynamical variable}
\label{sec:inv}

\vspace{0.5cm}

We study the gravity action (\ref{ActionGen}) in FLRW metric (\ref{FLRW})
$$
ds^{2} =N^{2}(\tilde{t})\,d\tilde{t}^{2}-a^{2}(\tilde{t})\,d\vec{x}^{2}\,,\ \ \ \ \ N(\tilde{t})>0\,,\ \ a(\tilde{t})>0\,.
$$
  In this case, the part of the action quadratic in curvature (\ref{A2}) equals to $\int\,d\tilde{t}\,\sqrt{-\mathcal{G}}\,R^{2}$ up to the boundary terms. For simplicity, we assume these boundary terms to be equal to zero. That is, we consider the model with the action
 $$
 A =A_{0}+A_{1}+A_{2}
$$
\begin{equation}
   \label{Aflrw}
 =\Lambda\int\,d\tilde{t}\,\sqrt{-\mathcal{G}}
-\frac{\kappa}{6}\,\int\,d\tilde{t}\,\sqrt{-\mathcal{G}}\,R+\frac{\lambda^{2}}{72}\int\,d\tilde{t}\,\sqrt{-\mathcal{G}}\,R^{2}\,.
\end{equation}

Now  the general coordinate invariance of the action is reduced to its invariance under the group of reparametrizations of the time coordinate. We suppose it to be the group of diffeomorphisms\footnote{The consequences of the invariance under its subgroup, $SL(2,\,\mathbf{R})$ group, for quantum cosmology was studied in \cite{(Livine1)}, \cite{(Livine2)}.} of
 the real semiaxis including zero $ Diff \left( \mathbf{R}^{+}\right)$.

The two coordinate systems are the most popular. They are the so-called cosmological coordinate system where $N(t)=1\,,$ and the so-called conformal coordinate system where $N(\tau)=a(\tau)\,,$ with cosmological time $t$ and conformal time $\tau$ being the time variable in the corresponding coordinate system.

Consider  diffeomorphisms  $\varphi \in Diff \left( \mathbf{R}^{+}\right)\,.$
We define the action of the diffeomorphism $\varphi$ on the functions $N(\tilde{t})$ and $a(\tilde{t})$ as follows:
\begin{equation}
   \label{fiNa}
\varphi\,N(\tilde{t})=\left(\varphi^{-1}(\tilde{t}) \right)'\,N\left(\varphi^{-1}(\tilde{t}) \right)\,;\ \ \ \ \ \varphi\,a(\tilde{t})=a\left(\varphi^{-1}(\tilde{t}) \right)\,.
\end{equation}

Instead of the laps and the scale factors, it is convenient to use the functions $f(\tilde{t})$ and $h(\tilde{t})$ defined by the following equations:
\begin{equation}
   \label{f}
\left(f^{-1}(\tilde{t}) \right)'=\frac{N(\tilde{t})}{a(\tilde{t})}\,,\ \ \ \ \ \ \ f^{-1}(\tilde{t}) =\int\limits_{0}^{\tilde{t}}\,\frac{N(\tilde{t}_{1})}{a(\tilde{t}_{1})}\,d\tilde{t}_{1}\,,\ \ \ f^{-1}(0) =0\,;
\end{equation}
\begin{equation}
   \label{h}
h'(\tilde{t}) =N(\tilde{t})\,,\ \ \ \ \ \ \ h(\tilde{t}) =\int\limits_{0}^{\tilde{t}}\,N(\tilde{t}_{1})\,d\tilde{t}_{1}\,,\ \ \ h(0) =0\,,
\end{equation}
with the transformation rules\footnote{We denote the composition of the functions by "$\circ$ product". } under the action of the diffeomorphism $\varphi$
\begin{equation}
   \label{varphif-1}
(\varphi f)^{-1}(\tilde{t}) =\int\limits_{0}^{\tilde{t}}\,\frac{\varphi N(\tilde{t}_{1})}{\varphi a(\tilde{t}_{1})}\,d\tilde{t}_{1}=\int\limits_{0}^{\varphi^{-1}(\tilde{t})}\,\frac{N(\tilde{t}_{2})}{a(\tilde{t}_{2})}\,d\tilde{t}_{2}
=f^{-1}\left(\varphi^{-1}(\tilde{t})\right)\equiv
\left(f^{-1}\circ \varphi^{-1}\right)(\tilde{t}) \,;
\end{equation}
\begin{equation}
   \label{varphih}
(\varphi h)(\tilde{t}) =\int\limits_{0}^{\tilde{t}}\,\varphi N(\tilde{t}_{1})\,d\tilde{t}_{1}=\int\limits_{0}^{\varphi^{-1}(\tilde{t})}\,N(\tilde{t}_{2})\,d\tilde{t}_{2}=h\left(\varphi^{-1}(\tilde{t})\right)\equiv
\left(h\circ \varphi^{-1}\right)(\tilde{t}) \,.
\end{equation}

From (\ref{varphif-1}),  the equation
\begin{equation}
   \label{varphif}
\varphi f =
\varphi \circ f
\end{equation}
follows immediately.

The invariance of the action under the group of diffeomorphisms looks like
\begin{equation}
   \label{invA}
A(f,\,h)=A\left(\varphi f,\,\varphi h\right)=A\left(\varphi \circ f,\,h\circ\varphi ^{-1}\right)\,.
\end{equation}

Note that the function
\begin{equation}
   \label{barg}
 g= h\circ f
\end{equation}
 is invariant under the diffeomorphisms $\varphi\,:$
$$
 g= h\circ \varphi^{-1} \circ \varphi \circ f=h\circ f\,.
$$

Choosing the diffeomorphism $\varphi$ to be
\begin{equation}
   \label{varphi=h}
\varphi  = h\,,
\end{equation}
we obtain the cosmological coordinate system
\begin{equation}
   \label{acosmol}
N(t)=1\,,\ \ \ \ \ a(t)=g'\left(g^{-1}(t) \right)\,,
\end{equation}
\begin{equation}
   \label{cosmolmetric}
ds^{2} =dt^{2}-\left(g'\left(g^{-1}(t) \right)\right)^{2}\,d\vec{x}^{2}\,.
\end{equation}

And if we choose,
\begin{equation}
   \label{varphi=f}
\varphi  = f^{-1}\,,
\end{equation}
we have the conformal coordinate system
\begin{equation}
   \label{confmetric}
ds^{2} =\left(g'(\tau)\right)^{2}\,\left[d\tau^{2}-d\vec{x}^{2}\right]\,,\ \ \ \ \ N(\tau)=a(\tau)=g'(\tau)\,.
\end{equation}

Therefore, the functions $f(\tilde{t})$ and $h(\tilde{t})$ have the transparent physical meanings.
$h$ is the reparametrization of the time $\tilde{t}$ into the cosmological time $t\,,$ while $f^{-1}$ stands for
the reparametrization of the time $\tilde{t}$ into the conformal time $\tau\,.$

The function
$$
 g(\tau)=\left( h\circ f\right)(\tau)=h\left(f(\tau)\right)
$$
transforms the conformal coordinate system into the cosmological one
\begin{equation}
   \label{taut}
t=g(\tau)\,,\ \ \ \ \ \ \ \ \ \  \tau=g^{-1}(t)\,.
\end{equation}

The invariance of the action manifests itself in its dependence on the only invariant function $g$
\begin{equation}
   \label{AgRight}
A=A\left(g\right)=A_{0}\left(g\right)+A_{1}\left(g\right)+A_{2}\left(g\right)\,,
\end{equation}
with the explicit form
\begin{equation}
   \label{Ag0}
A_{0}\left( g\right)=\Lambda\,\int\,\left(g'(\tau) \right)^{4}\,d\tau\,,
\end{equation}
\begin{equation}
   \label{Ag1}
A_{1}\left( g\right)=-\kappa\,\int\,\left[\left(g''(\tau) \right)^{2}-\left(g''(\tau)\,g'(\tau) \right)'\,\right]\,d\tau\,,
\end{equation}
and
\begin{equation}
   \label{Ag2}
A_{2}\left( g\right)=\frac{\lambda^{2}}{2}\,\int\,\left(\frac{g'''(\tau)}{ g'(\tau)} \right)^{2}\,d\tau\,.
\end{equation}

Thus, every four-dimensional space-time (and the corresponding space-time FLRW metric) is determined by its proper function $g\,,$ and vice versa, every function $g$ determines the particular four-dimensional space-time. Averaging over the set of functions $g$ means the averaging over the set of possible four-dimensional spaces.

Note that the anomaly induced effective action for the conformal factor $\Phi(x)\,,$
$$
\mathcal{G}_{\mu\nu}(x)=\exp\{2\Phi(x)\}\,\mathcal{\bar{G}}_{\mu\nu}(x)\,,
$$
proposed in the seminal papers \cite{(Antoniadis0)},  \cite{(Antoniadis2)}, is reduced in the case of FLRW metric exactly to
(\ref{Ag0}) - (\ref{Ag2}) with
\begin{equation}
   \label{ConfFact}
\Phi=\log\,g'\,.
\end{equation}

In the next two sections where we study path integrals in the theory, we deal with the solution of the Euler-Lagrange equation of the type (see (\ref{gcl}) in appendix A):
$$
\left(g_{0}^{(1)}\right)'(\tau)=\sigma\,\tau\,.
$$

\vspace{0.5cm}

\section{Path integrals measure}
\label{sec:measure}

\vspace{0.5cm}

Now we represent path integrals in the theory
\begin{equation}
   \label{PIA}
\int\,F(g)\,\exp\{-A(g) \}\,dg
\end{equation}
as the integrals of the form
\begin{equation}
   \label{PImeasure}
\int\,F(g)\,\exp\{ -A_{0}(g) -A_{1}(g)\}\,\mu_{\lambda}(dg)\,,
\end{equation}
over the functional measure
\begin{equation}
   \label{measure}
\mu_{\lambda}(dg)=\exp\{ -A_{2}(g) \}\ dg=\exp\left\{-\frac{\lambda^{2}}{2}\,\int\,\left(\frac{g'''(\tau)}{ g'(\tau)} \right)^{2}\,d\tau \right\}\ dg\,.
\end{equation}

 If we substitute
\begin{equation}
   \label{q}
q(\tau)=\frac{g''(\tau)}{g'(\tau)}\,,
\end{equation}
we can rewrite the integral in the exponent in the measure density (\ref{measure}) as
$$
-\frac{\lambda^{2}}{2}\,\int\,\left(\frac{g'''(\tau)}{ g'(\tau)} \right)^{2}\,d\tau
$$
\begin{equation}
   \label{qpmeasure}
=-\frac{\lambda^{2}}{2}\,\int\,\left[(q'(\tau))^{2}+2q'(\tau)q^{2}(\tau)+q^{4}(\tau) \right]\,d\tau=-\frac{\lambda^{2}}{2}\,\int\,(p'(\tau))^{2}\,d\tau\,,
\end{equation}
where $p$ is given by the nonlinear nonlocal substitution
\begin{equation}
   \label{p}
p(\tau)=q(\tau)+\int\limits_{0}^{\tau}q^{2}(\tau_{1})\,d\tau_{1}\,.
\end{equation}

The nonlinear nonlocal substitutions in the Wiener measure were studied in \cite{(BSh1)}, \cite{(BSh2)}. In particular, it was demonstrated that the space of integration  over the variable $q$ is different from that over the variable $p\,.$

While the paths  $p(\tau)$ form the space of all continuous  functions on the interval $[0,\,T]\,,$ the paths $q(\tau)$ are continuous almost at all points of the interval but may have singularities of the form
$$
q(\tau)\sim \frac{1}{\tau-\tau_{j}^{\ast}}
$$
at a finite number of points of the finite interval $[0,\,T]\,.$

The result follows directly from the representation
\begin{equation}
   \label{ksi}
q(\tau)=p(\tau)-\xi(\tau)\,,
\end{equation}
where $\xi(\tau)$ is the solution of the equation
\begin{equation}
   \label{ksi1}
\xi'(\tau)=\left(p(\tau)-\xi(\tau)\right)^{2}\,.
\end{equation}
So, there is a point $\tau_{+}$ where $\xi(\tau_{+})>0\,.$

For the function
\begin{equation}
   \label{eta}
\eta(\tau)=\frac{1}{\xi(\tau)}
\end{equation}
 the differential equation has the form
\begin{equation}
   \label{eta1}
\eta'(\tau)=-1+2p(\tau)\eta(\tau)-p^{2}(\tau)\eta^{2}(\tau)\,.
\end{equation}
As $\ \eta'(\tau)<0 \ $ and $\eta(\tau_{+})>0\,,$ there can be a point $ \tau^{\ast}$ where  $\eta\left(\tau^{\ast}\right)=0\,.$
In the vicinity of this point,
$$
\eta(\tau)\sim -(\tau-\tau^{\ast})\,,
$$
and
$$
\xi(\tau)\sim -\,\frac{1}{\tau-\tau_{j}^{\ast}}\,.
$$
The number of singular points and their positions are different for different paths $p(\tau)\,.$

Now we express the function $g(\tau)$ in terms of $p(\tau)\,.$
Here, we
choose the initial conditions
\begin{equation}
   \label{initcond}
g(0)=0\,,\ \ \ \ \ g'(0)=0\,,\ \ \ \ \ g''(0)=\sigma\,,
\end{equation}
consistent with the classical solution (\ref{gcl}) $\left(g_{0}^{(1)}\right)'(\tau)=\sigma\,\tau $.

At the interval $\left[\bar{\tau}_{1}\,,\,\bar{\tau}_{2}\right]$ without singular points $\tau^{\ast} \,$ in it, the solution of (\ref{q}) looks like
$$
g(\tau)=g(\bar{\tau}_{1})+g'(\bar{\tau}_{1})\,\int\limits_{\bar{\tau}_{1}}^{\tau}\exp\left
\{\int\limits_{\bar{\tau}_{1}}^{\bar{\tau}}q(\tau_{1})d\tau_{1}\right\}\,d\bar{\tau}
$$
\begin{equation}
   \label{gq}
=g(\bar{\tau}_{1})+g'(\bar{\tau}_{1})\,\int\limits_{\bar{\tau}_{1}}^{\tau}\exp\left
\{\int\limits_{\bar{\tau}_{1}}^{\bar{\tau}}\left[p(\tau_{1})-\xi(\tau_{1})\right]d\tau_{1}\right\}\,d\bar{\tau}\,.
\end{equation}

In case a singular point $\tau^{\ast}$ is in the region of integration $\tau^{\ast} \in \left[\bar{\tau}_{2}\,,\,\bar{\tau}_{3}\right] $,
the integral in the exponent can be written as\footnote{By virtue of the linear behavior of $\eta(\tau)$ in the vicinity of the singular point $\tau^{\ast}$, the integral $
\int\limits_{\bar{\tau}_{2}}^{\bar{\tau}}\frac{\eta'(\tau_{2})}{\eta(\tau_{2})}d\tau_{2}
$ can be defined in the sense of principal value. }
$$
\int\limits_{\bar{\tau}_{2}}^{\bar{\tau}}q(\tau_{2})d\tau_{2}=\int\limits_{\bar{\tau}_{2}}^{\bar{\tau}}p(\tau_{2})d\tau_{2}-
\int\limits_{\bar{\tau}_{2}}^{\bar{\tau}}\frac{1}{\eta(\tau_{2})}d\tau_{2}
$$
$$
=\int\limits_{\bar{\tau}_{2}}^{\bar{\tau}}p(\tau_{2})d\tau_{2}-
\int\limits_{\bar{\tau}_{2}}^{\bar{\tau}}\frac{-\eta'(\tau_{2})+2p(\tau_{2})\,\eta(\tau_{2})-p^{2}(\tau_{2})\,\eta^{2}(\tau_{2})}{\eta(\tau_{2})}
d\tau_{2}
$$
$$
=\ln\left|\eta(\bar{\tau}) \right|- \ln\left|\eta(\bar{\tau}_{2})\right|-
\int\limits_{\bar{\tau}_{2}}^{\bar{\tau}}\,p(\tau_{2})\,\left[1-p(\tau_{2})\,\eta(\tau_{2})\right]
d\tau_{2},.
$$

Therefore, at the interval $\left[\bar{\tau}_{2}\,,\,\bar{\tau}_{3}\right]$ with a singular point $\tau^{\ast}$ in it, the function $g(\tau)$ is
expressed in terms of $p$ and $\eta$ by the equation
\begin{equation}
   \label{gpeta}
g(\tau)=g(\bar{\tau}_{2})+\frac{g'(\bar{\tau}_{2})}{\left|\eta(\bar{\tau}_{2})\right| }\,\int\limits_{\bar{\tau}_{2}}^{\tau}\,\left|\eta(\bar{\tau}) \right| \,\exp\left
\{\int\limits_{\bar{\tau}_{2}}^{\bar{\tau}}\,p(\tau_{2})\,\left[1-p(\tau_{2})\,\eta(\tau_{2})\right]
d\tau_{2}\right\}\,d\bar{\tau}
\,.
\end{equation}

Then we consider the interval $\left[\bar{\tau}_{3}\,,\,\bar{\tau}_{4}\right]$ without  singular points in it, and use the equation (\ref{gq}).
We continue the process dividing the region of integration into the proper intervals and alternating the functions $\xi(\tau)\,,\ \eta (\tau)$
used in (\ref{gq}) and (\ref{gpeta}), and sewing together the values at the ends of the intervals:
\begin{equation}
   \label{sewing}
\eta(\bar{\tau}_{1})=\frac{1}{\xi(\bar{\tau}_{1})}\,,\ \ \ \ \ \xi(\bar{\tau}_{2})=\frac{1}{\eta(\bar{\tau}_{2})}\,,\ \ \ \ \ \eta(\bar{\tau}_{3})=\frac{1}{\xi(\bar{\tau}_{3})}\,,\ \ \ldots\,.
\end{equation}

In this case, $\eta(0)=0\,,$ and the function $g(\tau)$ is written as
\begin{equation}
   \label{g0peta}
g(\tau)=\sigma\,\int\limits_{0}^{\tau}\,\left|\eta(\bar{\tau}) \right| \,\exp\left
\{\int\limits_{0}^{\bar{\tau}}\,p(\tau_{1})\,\left[1-p(\tau_{1})\,\eta(\tau_{1})\right]
d\tau_{1}\right\}\,d\bar{\tau}
\,.
\end{equation}
at the interval $\left[0\,,\,\bar{\tau}_{1}\right]\,.$

Thus we establish the one-to-one correspondence between the function $g(\tau)$ and the Wiener variable $p(\tau)\,.$ It should be stressed that,
in spite of the singular character of the functions $\xi(\tau)\,,\ \eta (\tau)\,,$ the function $g(\tau)$  is continuous.

Now the measure $\mu_{\lambda}(dg)$ written in terms of $p(\tau)$ is the Wiener measure $w_{\frac{1}{\lambda}}(dp)\,.$

\vspace{0.5cm}

\section{First order perturbative correction to the scale factor}
\label{sec:pertcorr}

\vspace{0.5cm}

According to (\ref{acosmol}), the value of the scale factor at the moment of the cosmological time $t$ is the functional of $g$
$$
 a_{g}(t)=F(g)=g'\left(g^{-1}(t) \right)\,.
$$

For the classical solution
$$
g'_{cl}(\tau)=\left(g_{0}^{(1)}\right)'(\tau)=\sigma\,\tau\,,\ \ \ \ \ g_{cl}(\tau)=\frac{1}{2}\sigma\,\tau^{2}\,,\ \ \ \ \  g^{-1}_{cl}(t)=\sqrt{\frac{2\,t}{\sigma}}\,,
$$
the classical scale factor is equal to
$$
a_{cl}(t)=g'_{cl}\left(g_{cl}^{-1}(t) \right)=\sqrt{2\sigma\,t}\,.
$$

Now we define the scale factor averaged over the set of functions $g$ as
$$
<a(t)>_{g}=\mathcal{Z}^{-1}\,\int\,g'\left(g^{-1}(t) \right)
$$
\begin{equation}
   \label{aaverage}
\times\exp\left\{ \int\limits_{0}^{g^{-1}(t)}\left[-\Lambda\,\left(g'(\tau_{1}) \right)^{4}+\kappa\,\left(g''(\tau_{1}) \right)^{2}\right]\,d\tau_{1}-\kappa g''(g^{-1}(t))\,g'(g^{-1}(t))  \right\}\mu_{\lambda}(dg)
\end{equation}
with the normalizing factor
$$
\mathcal{Z}=\int\exp\left\{ \int\limits_{0}^{g^{-1}(t)}\left[-\Lambda\,\left(g'(\tau_{1}) \right)^{4}+\kappa\,\left(g''(\tau_{1}) \right)^{2}\right]\,d\tau_{1}-\kappa g''(g^{-1}(t))\,g'(g^{-1}(t))  \right\}\mu_{\lambda}(dg)\,.
$$

Our ansatz consists in cutting off the upper limit in integrals for the action terms  $A_{0}+A_{1}\,.$ For the action term $A_{2}$ entering
the measure density, the cut off occurs automatically due to the properties of the (Wiener) measure.

Due to the ansatz, the present is not influenced by the future.

To calculate the first nontrivial perturbative correction to $a_{cl}$, we represent the function $g(\tau)$ as (\ref{g0peta}), and expand the integrand in (\ref{aaverage}) up to the terms $O(p^{2})\,.$

Now the first factor in the integrand is written as
\begin{equation}
   \label{g1YX}
g'\left(g^{-1}(t) \right)=g'_{cl}\left(g_{cl}^{-1}(t) \right)+\sigma\,X+\sigma\,Y=\sqrt{2\sigma\,t}+\sigma\,X+\sigma\,Y\,,
\end{equation}
where $X$ contains the terms of the order $p$ and $Y$ contains the terms of the order $p^{2}\,.$
After the cancellation of the same terms in the nominator and the denominator, the factor $\exp\{-A_{0}-A_{1} \} $ gives
\begin{equation}
   \label{expE1}
 \left(1+E_{0}+E_{1}+O\left(p^{2} \right) \right)\,.
\end{equation}

(The explicit forms of $X\,,\ Y\,, \ E_{0}$ and $E_{1} $   are given in Appendix B.)

Thus
\begin{equation}
   \label{aWintegr}
<a(t)>=\sqrt{2\sigma\,t}+\sigma\,\int\ \left\{\,X\,\times\,\left(E_{0}+E_{1} \right)+Y\, \right\}\,w_{\frac{1}{\lambda}}(dp)\,.
\end{equation}

Then we change the order of the ordinary and path integration and use the following
simple rules for Wiener integration:
\begin{equation}
   \label{rules}
\int\,p(\tau_{1})\,w_{\frac{1}{\lambda}}(dp)=0\,,\ \ \ \ \ \ \
\int\,p(\tau_{1})\,p(\tau_{2})\,w_{\frac{1}{\lambda}}(dp)=\frac{1}{\lambda^{2}}\,\min\left\{ \tau_{1}\,,\,\tau_{2}\right\}\,,
\end{equation}
with the result
\begin{equation}
   \label{result}
<a(t)>=\sqrt{2\sigma\,t}\,\left\{1+\frac{1}{\lambda^{2}}\,\left[-\frac{59}{63}\left(\frac{2t}{\sigma} \right)^{\frac{3}{2}}
+\frac{11}{120}\,\kappa\,\left(2t\right)^{2}-\frac{1423}{2800}\,\Lambda\,\left(2t\right)^{4}\right]\right\}\,.
\end{equation}

\vspace{0.5cm}

\section{Conclusion}
\label{sec:concl}

\vspace{0.5cm}

In this paper, we study path integrals in the $R+R^{2}$ theory of gravity in the FLRW metric. The general coordinate invariance of the theory is reduced in this case to its invariance under the group of diffeomorphisms of the time coordinate.

We find the invariant of the group $g(\tau)$ that can be considered as the only dynamical variable in the theory.

In our approach,  Euclidean path integrals
are functional integrals not over the space of metrics $\mathcal{G}$, as they are commonly defined, but functional integrals
over the space of continuous functions $g(\tau)$ related to the conformal factor of the metric
$$
\int\,F(g)\,\exp\left\{-A (g)\right\}\,dg\,.
$$

Then we treat
path integrals in the theory  as integrals over the functional measure
$$
\mu(g)=\exp\left\{-A_{2} \right\}dg\,,
$$
where $A_{2}$ is the part of the action $A$ quadratic in $R\,.$ The rest part of the action in the exponent stands in the integrand as the "interaction" term.
In this sense, our approach is quite different from the commonly used ones. Usually the measure density contains the EH action $A_{1}$ in the exponent, while the other terms in the action are considered as the interaction.

Let us stress once more that, in our presentation of the theory, we deal with the only dynamical variable and with the well-defined path integrals
avoiding any ghosts problems.

For certain initial conditions $(\,g(0)=0\,)\,,$ we prove the measure  $\mu(g)$ to be equivalent to the Wiener measure and therefore not only rigorously define path integrals in quadratic gravity but also give the regular method of calculation. As an example of the approach, we calculate the first nontrivial perturbative  correction to the averaged scale factor $<a(t)>\,.$

From  (\ref{result}), one can see, in particular, the value of the time $t\,,$ when quantum corrections become significant. At the same time,
the experience gained in quantum field theory \cite{(BKShSV)} prevents us from considering first order perturbative results too seriously without
an analysis of the other terms. So, further studies in this direction are needed.

Another problem is to extend our method to the paths with initial conditions  different from (\ref{initcond}) and corresponding different types of space-time.

As it is discussed in section  \ref{sec:inv}, the effective action for the conformal factor of the metric
proposed in the seminal papers \cite{(Antoniadis0)},  \cite{(Antoniadis2)} is reduced to the action (\ref{Ag0}) - (\ref{Ag2}) in the case of FLRW metric.  The quantum properties of the general theory were also studied in the subsequent well known papers \cite{(Antoniadis3)} - \cite{(Antoniadis6)}. In particular, the detailed analysis of canonical quantization and the spectrum of physical states was performed there for the space-time of the form $\mathbf{R}\times S^{3}\,.$
Unfortunately,  right now we cannot  compare these results  of canonical quantization with ours obtained by path integration, because of the difference of the types of space-time considered.

So the elaboration of general methods for path integration over the measure (\ref{measure}) is highly desirable. Now it is in progress.
Note that the measure (\ref{measure}) is quasi-invariant under the action of the group of diffeomorphisms $Diff^{3}$ acting as a composition from the left
\begin{equation}
   \label{chi}
\chi\,g=\chi\,\circ\,g\,.
\end{equation}

That is,
\begin{equation}
   \label{muchi}
\mu_{\chi}(dg)\equiv \mu\left(d\,(\chi g)\right)=\mathcal{P}_{\chi}(g)\,\mu(dg)\,,
\end{equation}
where $\mathcal{P}_{\chi}(g)$ is the Radon-Nikodim derivative
$$
\mathcal{P}_{\chi}(g)=\exp\left\{-\frac{\lambda^{2}}{2}\,\int\,\left[\left(\frac{\chi'''(g)}{ \chi'(g)}\right)^{2}\left(g'(\tau) \right)^{4} +
6\frac{\chi'''(g)\,\chi''(g)}{\left(\chi'(g)\right)^{2}}\left(g'(\tau) \right)^{2}g''(\tau)\right.\right.
$$
\begin{equation}
   \label{radon}
\left.\left.+2\frac{\chi'''(g)}{\chi'(g)}\,g'''(\tau)\,g'(t)+9\left(\frac{\chi''(g)}{\chi'(g)}\right)^{2}\left(g''(\tau) \right)^{2}
+6\frac{\chi''(g)}{\chi'(g)}\,\frac{g''(\tau)g'''(\tau)}{g'(\tau)}
\right]\,d\tau \right\}\,.
\end{equation}

In \cite{(BShExact)} - \cite{(BShCalc)}, we calculated some nontrivial Wiener and Schwarzian functional integrals using the
properties of the measures.
We believe that quasi-invariance of the measure (\ref{measure}) and the explicit form of the Radon-Nikodim derivative (\ref{radon}) will be helpful also in this case.

\vspace{0.5cm}

\section{Acknowledgements}

\vspace{0.5cm}

We thank A. S. Ivanov and V. V. Chistyakov for their help in calculating some results in the paper.
We are grateful to D. I. Kazakov, V. A. Rubakov, A. A. Tseytlin, I. L. Shapiro and M. A. Vasiliev for valuable comments.

\vspace{0.5cm}

\section{Appendix A}

\vspace{0.5cm}

The Euler-Lagrange equation
$$
\left[\frac{\partial}{\partial g'}-\frac{d}{d\tau}\,\frac{\partial}{\partial g''}+\frac{d^{2}}{d\tau^{2}}\,\frac{\partial}{\partial g'''} \right]\,\mathcal{L}=0
$$
for the action (\ref{AgRight}) gives
\begin{equation}
   \label{ELeq}
2\Lambda\,\left(g' \right)^{3}-\kappa\,g'''+\frac{\lambda^{2}}{2}\,\left[6\frac{\left(g''\right)^{2}\,g'''}{\left( g'\right)^{4}}
-3\frac{\left(g'''\right)^{2}}{\left( g'\right)^{3}}-4\frac{g''\,g^{(4)}}{\left( g'\right)^{3}}+\frac{g^{(5)}}{\left( g'\right)^{2}}\right]=0\,.
\end{equation}

For the solutions of the form
$$
g'(\tau)=\sigma\,\tau^{\alpha}\,,\ \ \ \ \ \ \ \sigma=const\,,
$$
the equation (\ref{ELeq}) turns into
\begin{equation}
   \label{ELalfa}
2\Lambda\,\sigma^{4}\,\tau^{3\alpha}-\kappa\,\sigma^{2}\,\alpha\,(\alpha-1)\,\tau^{\alpha-2}+3\lambda^{2}\,\alpha\,(\alpha-1)\,(\alpha+1)\,
\tau^{-\alpha-4}=0\,.
\end{equation}

Besides the trivial solution
$
 \ \alpha=0\,,\ \Lambda=0\,,
$
that corresponds to the flat space-time  $R=0\,,$
there are solutions of the following two types:

$
(1). \ \alpha=1\,,\ \ \ \ \ \Lambda=0\,,
$ and

$
(2). \ \alpha=-1\,,\ \ \ \ \ \ \sigma^{2}=\frac{\kappa}{\Lambda}\,.
$

In the first case, the last term in (\ref{ELalfa}) dominates at small $\tau$ in the vicinity of the solution. And the constant $\sigma$ is not determined by the equation.
\begin{equation}
   \label{gcl}
\left(g_{0}^{(1)}\right)'(\tau)=\sigma\,\tau\,,\ \ \ \ \ g_{0}^{(1)}(\tau)=\frac{1}{2}\sigma\,\tau^{2}\,,\ \ \ \ \ \tau=\sqrt{\frac{2t}{\sigma}}\,,
\end{equation}
and
\begin{equation}
   \label{Rcl}
R_{0}^{(1)}(t)=\frac{\left(g''(\sqrt{\frac{2t}{\sigma}} )\right)^{2}}{\left(g'(\sqrt{\frac{2t}{\sigma}}) \right)^{4}}=\frac{1}{4t^{2}}\,,\ \ \ \ \ \ \ \ a_{0}^{(1)}(t)=\sqrt{2\sigma\,t}\,.
\end{equation}

Thus it is the classical picture of the universe arising from a point and expanding with the slowdown.

The classical solution of the second type looks like
\begin{equation}
   \label{3clsol}
\left(g_{0}^{(2)}\right)'(\tau)=\sigma\,\tau^{-1}\,.
\end{equation}

In this case, all the terms of the equation (\ref{ELalfa}) have approximately the same asymptotic behavior $(\sim\tau^{-3})$ at small $\tau$
in the vicinity of the solution.

It is the de Sitter space with $R_{0}^{(2)}=\frac{\Lambda}{\kappa}\,. $

The conformal time $\tau$ is related to the cosmological time $t$ by the equation
\begin{equation}
   \label{3tclsol}
t=g_{0}^{(2)}(\tau)=\sigma\,\int\limits_{\tau_{0}}^{\tau}\tau^{-1}=\sigma\,\ln\left|\frac{\tau}{\tau_{0}}\right|.
\end{equation}
and the scale factor looks like
\begin{equation}
   \label{a3clsol}
a_{0}^{(2)}(t)=\left(g_{0}^{(2)}\right)'\left(\,\left(g_{0}^{(2)}\right)^{-1}(t)\,\right)=\sigma\,\left(\tau_{0}\,\exp\{\frac{t}{\sigma} \}\right)^{-1}=\frac{\sigma}{\tau_{0}}\,\exp\{-\frac{t}{\sigma} \}\,.
\end{equation}

For positive $\tau$ (and $\sigma=\sqrt{\frac{\kappa}{\Lambda}}\,,$  $\tau_{0}>0)\,,$ the scale factor decreases exponentially.

However for
$
-\infty<\tau_{0} <\tau<0\,,\  \sigma=-\sqrt{\frac{\kappa}{\Lambda}}\,,
$
there is the exponential growth
\begin{equation}
   \label{a3growth}
a_{0}^{(2)}(t)=\left|\frac{\sigma}{\tau_{0}}\right|\,\exp\{\frac{t}{|\sigma|} \}\,.
\end{equation}

Note that $t>0\,$ in both cases.

\vspace{0.5cm}

\section{Appendix B}

\vspace{0.5cm}

Here, we present the explicit forms of the terms  $X\,,\ Y$ and $E $ entering (\ref{g1YX}), (\ref{expE1}) and (\ref{aWintegr}).
\begin{equation}
   \label{X}
X=- \sqrt{\frac{2\,t}{\sigma}}\,J_{1}+2J_{2}-2\left(\sqrt{\frac{2\,t}{\sigma}}\right)^{-1} J_{3}+ \left(\sqrt{\frac{2\,t}{\sigma}}\right)^{-1} J_{4}\,,
\end{equation}
$$
Y=p\left(\sqrt{\frac{2\,t}{\sigma}} \right)\,J_{4}-2p\left( \sqrt{\frac{2\,t}{\sigma}}\right)\,J_{3}-\sqrt{\frac{2\,t}{\sigma}}\,  J_{5}+J_{6}-4J_{7}+\frac{1}{2}\sqrt{\frac{2\,t}{\sigma}}\,  J_{1}^{2}-2J_{1}J_{2}
$$
$$
-\frac{1}{2} \left(\sqrt{\frac{2\,t}{\sigma}}\right)^{-1}   J_{8}+2 \left(\sqrt{\frac{2\,t}{\sigma}}\right)^{-1}   J_{9}+4  \left(\sqrt{\frac{2\,t}{\sigma}}\right)^{-1}   J_{10}-  \left(\sqrt{\frac{2\,t}{\sigma}}\right)^{-1}  J_{11}
$$
$$
+ \left(\sqrt{\frac{2\,t}{\sigma}}\right)^{-1}   J_{12}+4\left(\sqrt{\frac{2\,t}{\sigma}}\right)^{-2}    J_{2}J_{3}-2    \left(\sqrt{\frac{2\,t}{\sigma}}\right)^{-2}J_{2}J_{4}
$$
\begin{equation}
   \label{Y}
-2 \left(\sqrt{\frac{2\,t}{\sigma}}\right)^{-3}   J_{3}J_{3}-\frac{1}{2}  \left(\sqrt{\frac{2\,t}{\sigma}}\right)^{-3}  J_{4}J_{4}+2 \left(\sqrt{\frac{2\,t}{\sigma}}\right)^{-3}    J_{3}J_{4}
\end{equation}
\begin{equation}
   \label{E0}
E_{0}=-\Lambda\sigma^{4}\,\left[8J_{15}-4J_{16}-2 \left(\frac{2\,t}{\sigma}\right)^{\frac{3}{2}}\,  J_{14}+ \left(\frac{2\,t}{\sigma}\right)^{2}\,  J_{4}\right]\,,
\end{equation}
\begin{equation}
   \label{E1}
E_{1}=\kappa\sigma^{2}\,\left[-J_{13}-2J_{14}+2 \sqrt{\frac{2\,t}{\sigma}}\,  J_{1} \right]\,,
\end{equation}
where we denote
$$
J_{1}=\sqrt{\frac{2\,t}{\sigma}}\,\int\limits_{0}^{\sqrt{\frac{2\,t}{\sigma}}}d\tau_{1}\,p(\tau_{1})\,,\ \ \ \ \
J_{2}=\int\limits_{0}^{\sqrt{\frac{2\,t}{\sigma}}}d\tau_{1}\,\tau_{1}\,p(\tau_{1})\,,
$$
$$
J_{3}=\int\limits_{0}^{\sqrt{\frac{2\,t}{\sigma}}}d\tau_{1}\,\int\limits_{0}^{\tau_{1}}d\tau_{2}\,\tau_{2}\,p(\tau_{2})\,,\ \ \ \ \
J_{4}\int\limits_{0}^{\sqrt{\frac{2\,t}{\sigma}}}d\tau_{1}\,\tau_{1}\,\int\limits_{0}^{\tau_{1}}d\tau_{2}\,p(\tau_{2}) \,,
$$
$$
J_{5}=\int\limits_{0}^{\sqrt{\frac{2\,t}{\sigma}}}d\tau_{1}\,\tau_{1}\,p^{2}(\tau_{1})\,,\ \ \ \ \
J_{6}=\int\limits_{0}^{\sqrt{\frac{2\,t}{\sigma}}}d\tau_{1}\,\tau^{2}_{1}\,p^{2}(\tau_{1})\,,
$$
$$
J_{7}\int\limits_{0}^{\sqrt{\frac{2\,t}{\sigma}}}d\tau_{1}\,p(\tau_{1})\,\int\limits_{0}^{\tau_{1}}d\tau_{2}\,\tau_{2}\,p(\tau_{2})\,,\ \ \ \ \ J_{8}=\int\limits_{0}^{\sqrt{\frac{2\,t}{\sigma}}}d\tau_{1}\,\tau_{1}\,\int\limits_{0}^{\tau_{1}}d\tau_{2}\,p(\tau_{2})
\,\int\limits_{0}^{\tau_{1}}d\tau_{3}\,p(\tau_{3})\,,
$$
$$
J_{9}=\int\limits_{0}^{\sqrt{\frac{2\,t}{\sigma}}}d\tau_{1}\,\int\limits_{0}^{\tau_{1}}d\tau_{2}\,\tau_{2}\,p(\tau_{2})
\,\int\limits_{0}^{\tau_{1}}d\tau_{3}\,p(\tau_{3})\,,
\ \ \ \ \
J_{10}
=\int\limits_{0}^{\sqrt{\frac{2\,t}{\sigma}}}d\tau_{1}\,\int\limits_{0}^{\tau_{1}}d\tau_{2}\,p(\tau_{2})
\,\int\limits_{0}^{\tau_{2}}d\tau_{3}\,\tau_{3}\,p(\tau_{3})\,,
$$
$$
J_{11}=\int\limits_{0}^{\sqrt{\frac{2\,t}{\sigma}}}d\tau_{1}\,\int\limits_{0}^{\tau_{1}}d\tau_{2}\,\tau^{2}_{2}\,p^{2}(\tau_{2})\,,\ \ \ \ \
J_{12}=\int\limits_{0}^{\sqrt{\frac{2\,t}{\sigma}}}d\tau_{1}\,\tau_{1}\,\int\limits_{0}^{\tau_{1}}d\tau_{2}\,\tau_{2}\,p^{2}(\tau_{2})\,,
$$
$$
J_{13}=\left(\sqrt{\frac{2\,t}{\sigma}}\right)^{2}\,p\left(\sqrt{\frac{2\,t}{\sigma}}\right)\,,\ \ \ \ \ \ \
J_{14}=\int\limits_{0}^{\sqrt{\frac{2\,t}{\sigma}}}d\tau_{1}\,\int\limits_{0}^{\tau_{1}}d\tau_{2}\,p(\tau_{2})\,,
$$
$$
J_{15}=\int\limits_{0}^{\sqrt{\frac{2\,t}{\sigma}}}d\tau_{1}\,\tau^{3}_{1}\,\,\int\limits_{0}^{\tau_{1}}d\tau_{2}\,\tau_{2}\,p(\tau_{2})\,,\ \ \ \ \
J_{16}=\int\limits_{0}^{\sqrt{\frac{2\,t}{\sigma}}}d\tau_{1}\,\tau_{1}^{4}\,\int\limits_{0}^{\tau_{1}}d\tau_{2}\,p(\tau_{2})\,.
$$
Here we write the integrals out in the form convenient to integrate according to the rules (\ref{rules}).

The expansion of the integrand up to the terms $\sim p^{n}\,\ (n>2)$ needed for higher orders is straightforward.

\end{document}